\documentclass[aps,prl,reprint,balancelastpage,nofootinbib,preprintnumbers,superscriptaddress]{revtex4-1}

\usepackage{slashed}
\usepackage{bm}
\usepackage{latexsym,amssymb,amsmath,float,url,mathrsfs}
\usepackage{latexsym}
\usepackage{graphicx}
\usepackage{epstopdf}
\usepackage{amsmath}
\usepackage{comment}
\usepackage{subfigure}
\usepackage{natbib}
\usepackage{hyperref}
\usepackage{pifont}
\usepackage{blindtext}
\usepackage{adjustbox}
\usepackage{multirow}
\usepackage{tabularx}
\usepackage[table]{xcolor}
\usepackage{orcidlink}
\usepackage{tikz}
\usetikzlibrary{tikzmark}
\usepackage{fontawesome}

\usepackage{blkarray}
\usepackage{graphicx}
\usepackage{amsfonts}
\usepackage{soul}
\usepackage{amssymb}
\usepackage{amsmath}
\usepackage{cancel}
\usepackage{tcolorbox}
\usepackage{tikz-feynman}
\usepackage{tcolorbox}

\usepackage{graphics,appendix,afterpage,makecell} 

\definecolor{oucrimsonred}{rgb}{0.6, 0.0, 0.0}
\definecolor{persianblue}{rgb}{0.11, 0.22, 0.73}
\definecolor{forestgreen}{rgb}{0.13,0.35,0.13}
\definecolor{lightgray}{rgb}{0.83, 0.83, 0.83}
 \hypersetup{colorlinks, citecolor=oucrimsonred, linkcolor=black, urlcolor=oucrimsonred}
\definecolor{cornellred}{rgb}{0.7, 0.11, 0.11}
\definecolor{navyblue}{rgb}{0.0, 0.0, 0.5}
\definecolor{amethyst}{rgb}{0.6, 0.4, 0.8}
\definecolor{yellow}{rgb}{1.0, 1.0, 0.0}
\definecolor{firebrick}{rgb}{0.7, 0.13, 0.13}
\definecolor{tangerineyellow}{rgb}{1.0, 0.8, 0.0}
\definecolor{deepfuchsia}{rgb}{0.76, 0.33, 0.76}
\definecolor{amber}{rgb}{1.0, 0.75, 0.0}
\definecolor{VioletRed4}{rgb}{0.55, 0.13, .32}
\definecolor{indiagreen}{rgb}{0.07, 0.53, 0.03}
\definecolor{VioletRed4}{rgb}{0.55, 0.13, .32}
\newcommand{\be}{\begin{equation}}
\newcommand{\ee}{\end{equation}}
\newcommand{\bea}{\begin{equation} \begin{aligned}}
\newcommand{\eea}{\end{aligned} \end{equation}}

\definecolor{oucrimsonred}{rgb}{0.6, 0.0, 0.0}
\newcommand\vertarrowbox[3][6ex]{%
  \begin{array}[t]{@{}c@{}} #2 \\
  \left\uparrow\vcenter{\hrule height #1}\right.\kern-\nulldelimiterspace\\
  \makebox[0pt]{\scriptsize#3}
  \end{array}%
}

\definecolor{verdechiaro}{rgb}{0.6,1,0.6}
\definecolor{giallochiaro}{rgb}{1,1,0.6}
\definecolor{bluscuro}{rgb}{0.15, 0.2, 0.9}
\definecolor{verdes}{rgb}{0.1, 0.5, 0.1}%
\definecolor{tangerineyellow}{rgb}{1.0, 0.8, 0.0}

\definecolor{americanrose}{rgb}{1.0, 0.01, 0.24}
\definecolor{cobalt}{rgb}{0.0, 0.28, 0.67}
\definecolor{brandeisblue}{rgb}{0.0, 0.44, 1.0}
\definecolor{mycolor}{rgb}{0.0, 0.0, 0.5}
\definecolor{oxfordblue}{rgb}{0.0, 0.13, 0.28}
\definecolor{azure}{rgb}{0.0, 0.5, 1.0}
\definecolor{turquoiseblue}{rgb}{0.0, 1.0, 0.94}
\newtcolorbox{mynewbox}[1]{colback=white!5!white,colframe=azure!75!black,fonttitle=\bfseries,title=#1}
\newtcolorbox{mybox}{colback=mycolor!5!white,colframe=azure!75!black}
\newtcolorbox{mynamedbox}[1]{colback=mycolor!5!white,colframe=azure!75!black,title=#1}
\definecolor{venetianred}{rgb}{0.78, 0.03, 0.08}
\newtcolorbox{mynamedbox1}[1]{colback=venetianred!5!white,colframe=venetianred!80!black,title=#1}
\newtcolorbox{mynamedbox2}[1]{colback=azure!5!white,colframe=azure!80!black,title=#1}

\definecolor{verdes}{rgb}{0.1, 0.5, 0.1}%
\definecolor{cornellred}{rgb}{0.7, 0.11, 0.11}

\definecolor{VioletRed4}{rgb}{0.55, 0.13, .32}

\hypersetup{
     colorlinks   = true,
     citecolor    = violet,
     urlcolor     = violet,
     linkcolor    = violet}

\definecolor{rossocorsa}{rgb}{0.83, 0.0, 0.0}

\def\lsim{\mathrel{\rlap{\lower4pt\hbox{\hskip0.5pt$\sim$}}
    \raise1pt\hbox{$<$}}}         
\def\gsim{\mathrel{\rlap{\lower4pt\hbox{\hskip0.5pt$\sim$}}
    \raise1pt\hbox{$>$}}}         

\usepackage[normalem]{ulem}

\begin{document}

\title[]{Can We Detect Deviations from  Einstein's Gravity
in Black Hole Ringdowns?}


\author{A. Kehagias\orcidlink{0000-0002-9552-9366}}
\affiliation{Physics Division, National Technical University of Athens, Athens, 15780, Greece}
\affiliation{Department of Theoretical Physics and Gravitational Wave Science Center,  \\
24 quai E. Ansermet, CH-1211 Geneva 4, Switzerland}

\author{A. Riotto\orcidlink{0000-0001-6948-0856}}
\affiliation{Department of Theoretical Physics and Gravitational Wave Science Center,  \\
24 quai E. Ansermet, CH-1211 Geneva 4, Switzerland}


\begin{abstract}
\noindent
The quasinormal mode spectrum of gravitational waves emitted during the black hole ringdown relaxation phase, following the merger of a black hole binary, is a crucial target of gravitational wave astronomy.
By considering causality constraints on the on-shell graviton three-point couplings within a weakly coupled gravity theory, we present  arguments indicating that  the contributions to the physics of linear and quadratic quasinormal   modes from higher derivative gravity theories are either negligible or vastly suppressed for Schwarzschild and Kerr black holes. Their spectrum and interactions  are
dictated solely by Einstein's gravity.

%
\end{abstract}
\maketitle


\noindent
{\it Introduction.}
Gravitational wave (GW) observations of merging black hole (BH) binaries~\cite{KAGRA:2021vkt,LIGOScientific:2021sio} give access to the most dynamical regime of strong gravity accessible with current experimental efforts.
These observations probe a plethora of physical properties of compact objects, ranging from tidal and spin-induced deformations, to horizon absorption and their oscillation spectra.
The oscillation spectrum of black holes has received special attention both experimentally and theoretically because of its phenomenological interest and because its interpretation is simple, at least when nuances due to the dynamical nature of the merger process are accounted for~\cite{Baibhav:2023clw,Buoninfante:2024oxl}.
The experimental program of ``black hole spectroscopy'' consists of measuring multiple black hole quasinormal modes (QNMs) from GW observations of the ringdown waves following a binary merger~\cite{Detweiler:1980gk,Dreyer:2003bv,Berti:2005ys}. The goal is both to improve the understanding of the astrophysical properties of these systems, and to use them as new physics laboratories~\cite{Berti:2009kk,Berti:2018vdi,Cardoso:2019rvt}.

%
Equally interesting, a debate on the domain of validity of linear perturbation theory in the description of black hole binary dynamics (see e.g.~\cite{Buoninfante:2024oxl} and references therein) triggered a community effort to systematically explore nonlinear effects, such as absorption-induced  variations in the mass and spin~\cite{Sberna:2021eui,Redondo-Yuste:2023ipg,Zhu:2024dyl,May:2024rrg,Capuano:2024qhv} and transient effects~\cite{Berti:2006wq,Lagos:2022otp,Albanesi:2023bgi}.
Particular attention was devoted to quadratic and higher-order QNMs~\cite{Gleiser:1995gx,Brizuela:2009qd,Ioka:2007ak,Nakano:2007cj,Pazos:2010xf,Loutrel:2020wbw,Ripley:2020xby}.
These additional families of QNM frequencies naturally arise from the nonlinear nature of Einstein's equations -- in particular from the trilinear graviton couplings --  and they were recently confidently extracted from nonlinear binary merger simulations~\cite{London:2014cma,Cheung:2022rbm,Mitman:2022qdl,Cheung:2023vki,Baibhav:2023clw,Khera:2023oyf}.
After these breakthroughs, multiple analytical and numerical techniques were rapidly deployed to characterize the amplitudes and the initial data dependence of these modes for Schwarzschild and Kerr black holes~\cite{Bucciotti:2023ets, Perrone:2023jzq, Kehagias:2023ctr, Redondo-Yuste:2023seq,Ma:2024qcv,Bucciotti:2024zyp,Bucciotti:2024jrv,Bourg:2024jme,Zhu:2024rej,Ma:2024qcv,Khera:2024yrk,Kehagias:2024sgh}. These predictions can now be readily used as additional ingredients in the construction of binary black hole waveform templates and 
the  nonlinear effects could potentially be observed in a small number of events with current ground-based detectors, and the detection prospects are even more promising for the upcoming LISA mission \cite{Yi:2024elj, Lagos:2024ekd}.

In parallel, considerable efforts were dedicated to the construction of black hole spectroscopy models including higher derivative corrections beyond Einstein's gravity.
One of the goals of this program is to enhance the sensitivity of new physics searches, by incorporating indications (e.g. isospectrality breaking) given by effective field theory expansions under minimal assumptions.
Such searches had to previously rely on agnostic ansätze~\cite{Maselli:2019mjd,Carullo:2021dui,Payne:2024yhk} or employ slow-rotation predictions~\cite{Pani:2012bp,Pani:2013pma,Pierini:2021jxd,Wagle:2021tam,Srivastava:2021imr,Cano:2021myl,Pierini:2022eim,Silva:2022srr,Maselli:2023khq}, inevitably less accurate especially for the typical medium-large rotation rate of observed black holes.
New developments in the last few years have led to highly accurate predictions of QNM frequencies that are fairly accurate also for the large spins of interest in GW applications, and therefore can be used for black hole spectroscopy~\cite{Li:2022pcy,Hussain:2022ins,Cano:2023tmv,Chung:2024ira,Blazquez-Salcedo:2024oek}.
%
%
Considering the significantly larger complexity of these calculations, matching the progress obtained in GR to construct templates including higher derivative contributions is a daunting task. It is important to call into question the validity and self-consistency of any beyond-GR correction before embarking in these calculations and testing their validity in experiments involving black hole mergers~\cite{Berti:2015itd,Cano:2019ore,Ripley:2019aqj,Serra:2022pzl,Corman:2022xqg,Cayuso:2023aht}.

Here we argue that the causal structure of these theories implies that their linear  and quadratic QNMs   are identical to the Einstein case, or heavily suppressed beyond any foreseeable experimental accuracy.
%
%
Our conclusions hinge on the well-known causality violation argument concerning the trilinear graviton coupling, which we now summarize.

\vskip 0.5cm
\noindent
{\it The causality argument.}
Our  argument is grounded in the findings from Ref.~\cite{Camanho:2014apa}.  The main premise is that trilinear graviton couplings arising from higher-order terms in the gravitational action in flat spacetime cannot be excessively large, otherwise they would cause causality violations and make the theory inconsistent.

We focus on weakly coupled gravity theories within a tree-level approximation. It is established that, at large distances, these theories should converge to Einstein's gravity. However, higher derivative corrections can emerge at intermediate energies. By intermediate energies we mean those that are sufficiently low for the theory to remain weakly coupled, yet high enough to be influenced by these higher derivative terms.

For definiteness, consider a gravity theory characterized by the action

\begin{equation}
\label{action}
  S=\frac{1}{16\pi G}\int{\rm d}^4 x\sqrt{-g}\left(R+\alpha_4R^3+\cdots\right), 
\end{equation}
where  $\alpha_4$ has dimensions of  (length)$^4$,  $R^3=
R_{\mu\nu\sigma\rho}R^{\sigma\rho\delta\tau}R_{\delta\tau}^{\,\,\,\,\,\,\mu\nu}$, and the dots indicate higher-order terms in the Riemann tensor. 
It is worth noting that in four dimensions, an $R^2$ term corresponds to a Gauss-Bonnet term, which is topological and thus  we do not include it.  We are then left with only the Einstein term and the $\alpha_4$ term (or higher-order terms \cite{AccettulliHuber:2020oou}).

These scales must be significantly larger than the Planck length to ensure an intermediate energy regime where higher derivative corrections are relevant while maintaining weak coupling.

A key point is that these corrections can modify the three-graviton vertex, potentially causing causality violations at short distances by inducing (Shapiro) time delays with incorrect signs in certain Gedanken experiments. This issue cannot be resolved unless one posits that the complete theory includes an infinite number of states extending to arbitrarily high spins.

The chosen Gedanken experiment in Ref.~\cite{Camanho:2014apa} involves high-energy scattering of a polarized “probe” graviton against a coherent state “target” composed of $N$ polarized gravitons. We select a regime where the full $S$-matrix approximates the $N$th power of an almost trivial elastic two-body $S$-matrix. In a suitable large-$N$ limit, the $S$-matrix exponentiates, resulting in an eikonal-like phase $2\delta(\sqrt{s},b)$ (with $s$ as the center-of-mass energy squared of the probe graviton and $b$ as the impact parameter), from which we can derive the time delay

\begin{equation}
\label{deltat}
    \Delta t=2\partial_{\sqrt{s}}\delta(\sqrt{s},b)=\Delta t_{EH}\left(1\pm c_4\frac{\alpha_4^4}{b^8}+\cdots\right),
\end{equation}
where $\Delta t_{EH}$ represents the Einstein-Hilbert contributions, and $c_2$ and $c_4$ are coefficients of order unity. The choice of sign depends on the relative orientations of the helicities of the probe and the target. This means we can always adjust the polarizations so that, for $b \ll \alpha_4^{1/4}$, the time delay has the opposite sign of the usual Einstein-Hilbert time delay, illustrating the potential causality violation.
Thus, one must either set the coefficients to zero or introduce an infinite number of degrees of freedom appearing at distances smaller than the cutoff length scale \cite{Camanho:2014apa}.

\begin{figure}[t]
    \centering
    \includegraphics[width=0.3\textwidth]{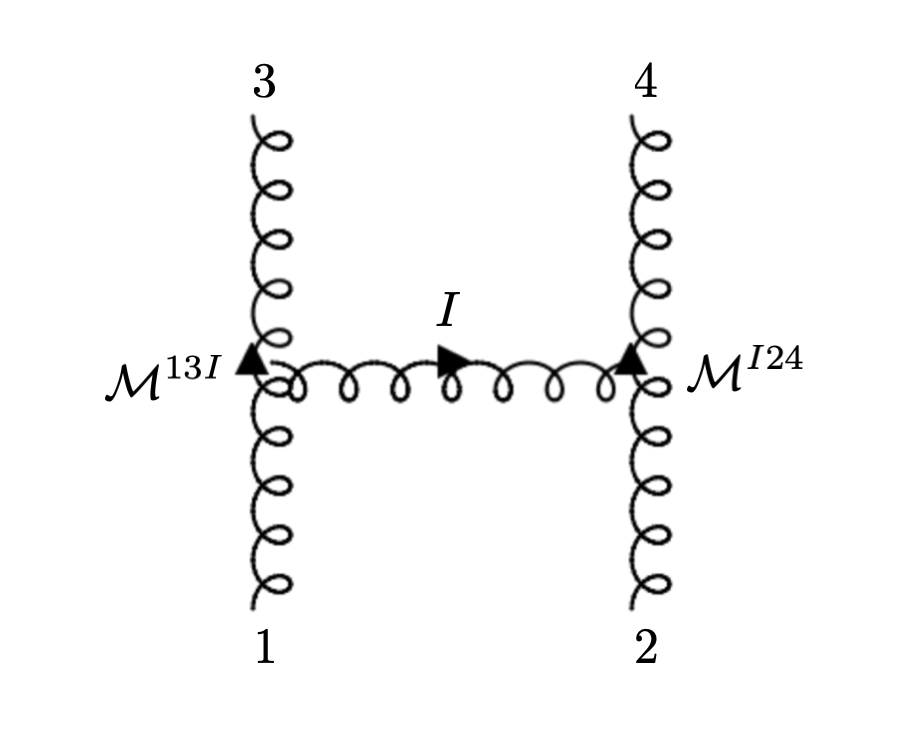}
    \caption{ Tree-level four-point graviton amplitude.}\label{fig:V1}
\end{figure}

An expert reader might well stop reading this section here. However, to delve deeper into this point,  which is the focus of our argument concerning the QNMs, let us examine the tree-level four-point amplitude ${\cal M}_4$ in flat spacetime for the graviton-graviton scattering in the $t$-channel, as depicted in Fig.~\ref{fig:V1}. This amplitude relies on the kinematic invariants generated by the four on-shell conserved momenta and polarization tensors of the external gravitons. As a tree-level amplitude, its singularities manifest only as poles in the Mandelstam variables. We can consider external momenta such that $s \gg t$, while keeping $Gs$ small enough for the theory to remain weakly coupled.

In impact parameter space, the phase shift for this interaction is expressed as

\begin{equation}
    \delta(\vec b,s)=\frac{1}{2s}\int\frac{{\rm d}^2\vec q}{(2\pi)^2}e^{i\vec q\cdot \vec b}{\cal M}_4(\vec q),
\end{equation}
where $s$ is the center-of-mass energy squared, $t \simeq -\vec q^2$, and $\vec b$ is the impact vector. For $t \ll s$, short-distance effects are negligible.
In this representation, for nonzero $\vec b$, we only receive contributions from diagrams that feature an on-shell particle in the $t$-channel~\cite{Camanho:2014apa}. The contribution from the massless graviton pole in the $t$-channel is

\begin{equation}
   \delta(\vec b,s)=\frac{1}{8\pi s} {\cal M}_3^{13I}(-i\partial_{\vec b}){\cal M}_3^{I24}(-i\partial_{\vec b}) \ln(L/b),
\end{equation}
where $L$ is an IR cutoff, $I$ represents the intermediate $t$-channel, and the derivatives $-i\partial_{\vec b}$ replace the momentum $\vec q$ in the three-point vertex ${\cal M}_3$. 

The inclusion of terms beyond the Einstein-Hilbert action may result in time advancements instead of delays when  $b \sim \alpha_4^{1/4}$. Notably, this can occur while the theory remains weakly coupled and the scattering is still within the eikonal regime. This phenomenon arises because gravity weakens with distance, counterbalancing the large energies involved. For instance, in the weakly coupled regime, with $Gs \ll 1$ and $\alpha_4 \gg G^2$, one finds

\begin{equation}
  \frac{\alpha_4^4}{b^8}\sim \alpha_2^4 t^4\ll \alpha_4^4 s^4\lsim \alpha_4^4G^{-8}(\gg 1),
\end{equation}
indicating that the parameter space permits causality violations. 

The  concept of causality applied here is relevant for scattering processes occurring in a flat and otherwise empty region of space and its implications will hold in any spacetime that contains a flat region with a radius much larger than $\alpha_4^{1/4}$, including of course our own spacetime.

If we demand that the theory remains causal, the coefficients of the higher derivative terms must either be identically zero, or new physics must enter through an infinite tower of extra massive particles with higher spins, akin to string theory. However, in this case, the energy cut-off for the theory must be sufficiently high to bypass all potential experimental constraints.

\vskip 0.5cm
\noindent
{\it The impact on the physics of  QNMs.}  
 The spectrum of QNMs is determined by the properties of the unstable circular null geodesics. These ``free'' vibrational modes can be understood as corresponding to null particles trapped at the unstable circular orbit, gradually escaping over time (see, for example, Ref.~\cite{Cardoso:2008bp}). For instance, for  a Schwarzschild black hole with mass $M$, the circular photon ring is located at $r = 3 GM$ and the spectrum of the QNMs is well approximated by \cite{Berti:2005ys}

 \begin{equation}
     \omega_{n\ell m}GM\simeq \frac{\ell}{3\sqrt{3}}-\frac{i}{3\sqrt{3}}\left(\frac{1}{2}+n\right),
 \end{equation}
  where $n=0$ is the fundamental mode and $(\ell,m)$ are the multipoles.

As a representative example, let us consider  the presence of an $R^3$  term in the gravity action. Broadly speaking and very schematically, in the Schwarzschild or Kerr spacetime, the Lagrangian will include quadratic and cubic  terms of the form

\begin{eqnarray}
   \mathscr{L}_{R+R^3}&\supset& (\partial h)^2+\alpha_4(\partial \overline{g})^4(\partial h)^2\nonumber\\
    &+&h(\partial h)^2+\alpha_4(\partial \overline{g})^4h(\partial h)^2+\cdots,
\end{eqnarray}
where $\partial \overline{g}\simeq 1/r_s$ is the derivative of the background geometry $\overline{g}$, $r_s=2G M$ is the Schwarzschild radius, and the metric has been expanded as $g=\overline{g}+h$. 

Imagine now two fundamental QNMs with $\ell=m=2$ produced during a ringdown phase following the merger of two black holes with masses $M\sim M_\odot$ and carrying energies $\omega_{022}\sim (M_\odot/M)$ km. Let them scatter when they  are  about to arrive at the Ligo and Vigo detectors, separated by distances much larger than a km, so that their exchanged momentum $ q$ in the $t$-channel is much smaller than  the center of mass energy $\sqrt{s}\sim \omega_{022}$.

The causality argument in flat spacetime applies  and establishes that the  couplings from higher derivative terms -- among which the trilinear graviton coupling sourcing the quadratic QNMs --  must be either zero, or that some new series of higher spin particles must intervene at scales smaller than the length scales in front of the higher derivative operators. 

In the first case, the spectrum and the nonlinearities of the QNMs are only sourced by the standard Einstein theory and the discussion would end here.

In the second case -- since during the ringdown phase the amplitude of the QNMs is such that $h\lsim G^{-1/2}$ -- any relative deviation for solar mass black holes from the QNM spectrum  and nonlinearities predicted by the standard Einstein's gravity is at most of the order of

\begin{eqnarray}
    \alpha_4\omega_{n\ell m}^4 \sim \frac{\alpha_4}{G^4 M^4} 
    \sim 10^{-32}\left(\frac{\alpha_4}{\mu{\rm m}^4}\right)\left(\frac{M_\odot}{M}\right)^4.
\end{eqnarray}
Indeed, gravity has been tested in tabletop experiments at scales as small as tens of microns, with results that align well with Newtonian theory. Given this, one should ideally prefer — if not insist — that any extension of general relativity be a theory valid down to the micron scale, accurately describing the same observations as general relativity.
We conclude again that, as long as weakly coupled gravity is considered, theoretical consistency dictates that the spectrum and the nonlinearities of the  QNMs can only arise from standard Einstein's gravity.

Of course, there is a third option:  one could envisage to increase the length scale up to the best measured mode and take

\begin{equation}
  \alpha_4^{1/4} \sim\omega_{022}^{-1} \sim \left(\frac{M}{M_\odot}\right)\,\,\,{\rm km}, 
\end{equation} 
in order to have a chance to detect some deviation from Einstein's gravity in the fundamental QNM. In such a case the  new higher spin particles would mediate a new force between all Standard
Model particles whose strength will be parametrically equal to the gravitational one. This  scenario is  clearly  not admissible in the absence of a realistic -- and consistent --   ultraviolet  completion capable of restoring causality without conflicting with our current understanding of gravity at short distances \cite{Endlich:2017tqa,Caron-Huot:2022ugt}. In this sense this  conclusion aligns well with those spelled out in Refs.~\cite{Serra:2022pzl,Caron-Huot:2022ugt}.

But let us be even more tenacious and  assume that some wizardry may soften gravity precisely on macroscopic scales of the order of the km, without 
pausing too long on why nature should have chosen this specific lenght scale.  If so, at what length scale exactly? Consider again a  black hole merger event which has  generated two fundamental QNMs with multipole $\ell\gg 2$ and let them scatter through a $t$-channel exchange of small momentum transfer away from the final black hole state, so that they can be safely  considered propagating in flat space time. The causality argument applies, obliging us to  push the cutoff length scale to be smaller than 

\begin{equation}
    \alpha_4^{1/4}\lsim \frac{1}{\omega_{0\ell m}}\sim \frac{2}{\omega_{022}\ell}\ll
   \frac{1}{\omega_{022}}
\end{equation}
and no effect will be visible in the fundamental $\ell=2$ mode, both at the liner and nonlinear level.

Our observations are not at odds  with those in Ref. \cite{Chen:2021bvg}, which argued that the apparent time advance is not necessarily problematic when the action (\ref{action}) is interpreted as an effective action and  one  requires to remain within the validity regime of the effective theory on a black hole background. For instance, the condition to ensure that the corrections to the linearized metric perturbation equations remain under control 
demands that the energy of the fundamental mode $\omega_{022}$ be significantly smaller than $\alpha_4^{-1/4}$. In turn,  this implies  again that no deviations from Einstein's theory would be observable in the QNM sector. Even more concerning, for sufficiently large multipole $\ell$, the resulting multipole inevitably violates causality for a specific choice of sign in the higher-dimensional operator coefficient \cite{deRham:2021bll}. Finally, the description of all QNM overtones cannot be made within the regime of validity of the effective field theory, a  feature which seems true of any extension to general relativity that introduces a new length as cutoff \cite{Silva:2024ffz}.

This raises the question of whether the effective field theory approach to gravity can ever describe modifications to Einstein’s theory that are not parametrically small. The answer seems to be  no: by employing a quantum concept of causality grounded in commutators and crossing symmetry, instead of a classical approach based on time advancements, if nature respects causality as we understand it, a higher spin particle must exist with a Compton wavelength at least as large as the length $\alpha_4^{1/4}$ and in any scenario where an effective field theory is valid at the distance $L$ (in our case $\sim \omega_{022}^{-1}$), the corrections are necessarily small $(\alpha_4^{1/4}/L \ll 1)$ \cite{Caron-Huot:2022ugt}. Again, this is consistent with the fact that  the absence of a consistent ultraviolet completion implies that the causal issues present in the low-energy effective action of a nonrenormalizable theory  remain unresolved even at high energy scales \cite{Hollowood:2016ryc}.


\vskip 0.5cm
\noindent
{\it Conclusions.} 
We have argued that  higher derivative gravity corrections to the spectrum and nonlinearities of QNMs must either identically vanish or must be heavily suppressed. Thus, any observable  QNM behavior must originate from the standard Einstein-Hilbert action. Extensions involving higher derivative terms would imply the presence of new particles or forces with significant observable effects, which are not supported by current experimental data at scales as small as tens of microns.

Our findings imply that, within the framework of weakly coupled gravity, the  QNM physics is  dictated solely by Einstein's gravity, affirming the robustness of general relativity at both the classical and perturbative quantum levels. 
If corrections to the properties of QNMs were to be detected, reconciling them with our arguments would require a drastic change and new physics beyond the current theoretical landscape.

\begin{acknowledgments}
\vspace{5pt}\noindent\emph{Acknowledgments.}
We thank E.~Berti and G. Carullo for useful discussions and feedback on the  V.~Cardoso, G. Carullo, K. Fransen, L.~Senatore and F.~Serra for useful discussions and comments on the manuscript.
%
%
A.R.  acknowledges support from the  Swiss National Science Foundation (project number CRSII5\_213497) and from   the Boninchi Foundation for the project ``Pblack holes in the Era of GW Astronomy''.
A.K. acknowledges support from the Swiss National Science Foundation (project number IZSEZ0 229414).
%

\end{acknowledgments}

\vskip 1cm
\noindent

\bibliography{Draft}

\end{document}